\def\year{2021}\relax
\documentclass[letterpaper]{article} 
\usepackage{aaai21}  
\usepackage{times}  
\usepackage{helvet} 
\usepackage{courier}  
\usepackage[hyphens]{url}  
\usepackage{graphicx} 
\usepackage{natbib}  
\usepackage{caption} 
\usepackage{hyperref}
\usepackage{url}            
\usepackage{booktabs}       
\usepackage{amsfonts}       
\usepackage{nicefrac}       
\usepackage{microtype}      
\usepackage{cite}
\usepackage{graphicx}
\usepackage{multirow}
\usepackage{xcolor}
\usepackage{makecell}
\usepackage{subcaption}
\usepackage{arydshln}
\usepackage{colortbl}
\usepackage{array}
\usepackage{color}
\usepackage{cleveref}
\usepackage{enumitem}
\usepackage{amsthm}
\usepackage{mathrsfs}
\usepackage{selectp}
\usepackage{xspace}
\usepackage[ruled,linesnumbered]{algorithm2e}
\definecolor{llg}{rgb}{0.9,0.9,0.9}
\title{Heuristic Semi-Supervised Learning for Graph Generation \\Inspired by Electoral College}
\author {

        Chen Li\textsuperscript{\rm 1},
        Xutan Peng\textsuperscript{\rm 2},
        Hao Peng\textsuperscript{\rm 1}, 
        Jianxin Li\textsuperscript{\rm 1},
        Lihong Wang\textsuperscript{\rm 3},
        Philip S. Yu\textsuperscript{\rm 4},
        Lifang He\textsuperscript{\rm 5}
        \\
}
\affiliations {
    \textsuperscript{\rm 1}Beihang University,
    \textsuperscript{\rm 2}University of Sheffield, \textsuperscript{\rm 3}CNCERT, \textsuperscript{\rm 4}University of Illinois at Chicago, \textsuperscript{\rm 5}Lehigh University\\
    \{lichen, penghao, lijx\}@act.buaa.edu.cn, 
    x.peng@shef.ac.uk,
    wlh@isc.org.cn,
    psyu@uic.edu,
    lih319@lehigh.edu
}

\begin{document}

\maketitle

\begin{abstract}
Recently, graph-based algorithms have drawn much attention because of their impressive success in semi-supervised setups.
For better model performance, previous studies learn to transform the topology of the input graph.
However, these works only focus on optimizing the original nodes and edges, leaving the direction of augmenting existing data unexplored.
In this paper, by simulating the generation process of graph signals, we propose a novel heuristic pre-processing technique, namely \textit{ELectoral COllege} (ELCO), which automatically expands new nodes and edges to refine the label similarity within a dense subgraph.
Substantially enlarging the original training set with high-quality generated labeled data, our framework can effectively benefit downstream models. 
To justify the generality and practicality of ELCO, we couple it with the popular Graph Convolution Network and Graph Attention Network to perform extensive evaluations on three standard datasets.
In all setups tested, our method boosts the average score of base models by a large margin of 4.7 points, as well as consistently outperforms the state-of-the-art.
We release our code and data on \url{https://github.com/RingBDStack/ELCO} to guarantee reproducibility.
\end{abstract}

\section{Introduction}\label{sec:intro}

Numerous real-world data can be represented as graphs, e.g., social networks~\citep{sn2018frai}, citation networks~\citep{data2016icml}, knowledge graphs~\citep{rgcn2018eswc}, and protein-interaction networks~\citep{protein2017nips}.
In many cases, large-scale annotated data is expensive to obtain.
The so-called graph-based Semi-Supervised Learning (SSL), which holds promise to bootstrap applications even with limited supervision, has therefore attracted increasing research interest.

Earlier works develop the classical regularization methods, which achieve SSL by smoothing feature representations or model predictions over local neighborhoods using explicit regularization schemes~\citep{lb2003icml,gf2013icml,mr2006imlr,ds2012nn}.
Although this direction has been well studied, a later thread of algorithms, namely graph convolution networks, have demonstrated state-of-the-art performance and drawn much attention~\citep{gcn2017iclr,gat2018iclr,cheby2016nips}.
By utilizing various aggregation strategies, these models selectively fuse the local features of a graph into the hidden representations of its target nodes.
To further perform downstream tasks, the hidden layers are coupled with specific task layers~\citep{gcn2017iclr,gat2018iclr}.
One common characteristic of these two lines of models is that they both adopt the presence of smoothness within the graph structure as a basic assumption.

Recently, to better exploit annotated resources, some researchers propose to modify the topology of the input graph.
For instance, DropEdge~\citep{dropedge2020iclr} prevents excessive smoothing by simplifying edges (i.e., randomly dropping a certain number of edges in the given graph); \citet{mixhop2019icml} adjust the local distribution of nodes by repeatedly mixing neighborhoods at various scales; \citet{top2019ijcai} restructure the graph based on modularity, thus strengthening the intra-community connections but reducing the inter-community ones. However, to the best of our knowledge, all such methods are limited within handling the \textit{existing} graph topology.

\begin{figure}
	\centering
	\includegraphics[width=0.5\textwidth]{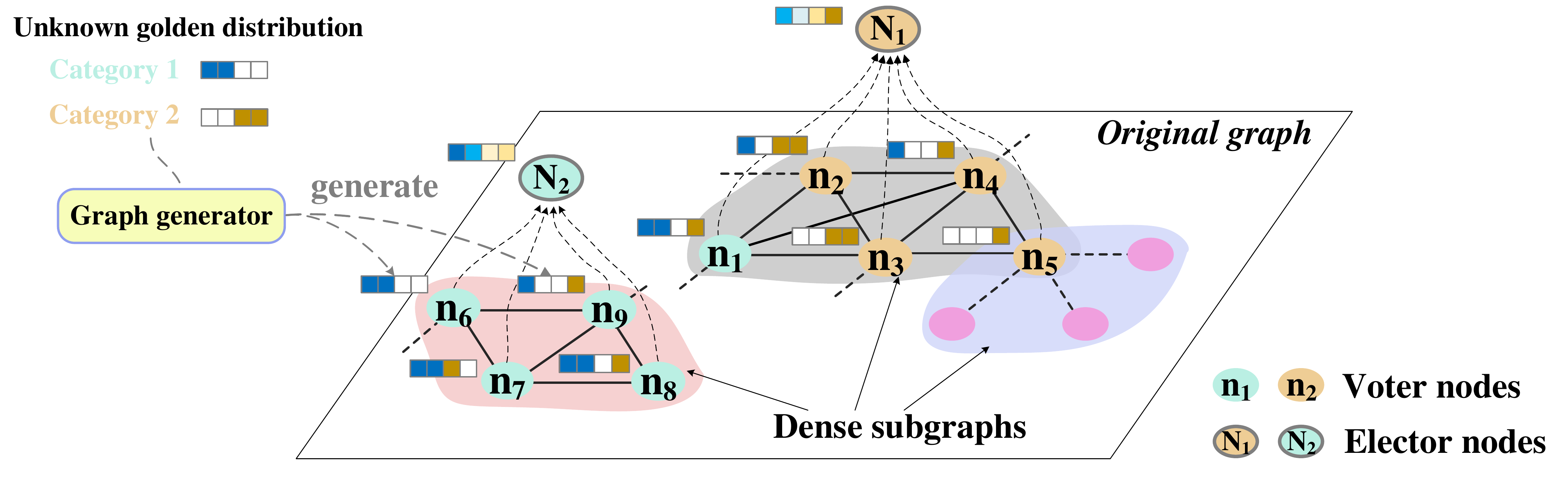}
	\caption{An illustrated example of ELCO. 
		$G_{1}\!=\!\{n_{1}, \cdots,n_{5}\}$ and $G_{2} = \{n_{6},\cdots,n_{9}\}$ are two dense subgraphs/clusters. 
		The color of node $n_{i}$ denotes its label $C_{n_{i}} \in \{1,\cdots,|C|\}$, and the upper left bar of each node shows the node attributes (following \citet{nodenoise2020arxiv}, they carry noise). 
		Labeled elector nodes $N_{1}$ and $N_{2}$ are obtained by aggregating node attributes and inheriting labels from their clusters. Obviously, the difference between attributes of elector nodes with different labels is more stable and bigger than voter nodes.}
	\label{fig:example}
\end{figure}

In this paper, we explore a novel research direction for the first time, which aims to expand the original graph by generating new nodes and edges.
More concretely, our \textit{ELectoral COllege} (ELCO) framework, which is inspired by the widely-known Electoral College system of the United States, first identifies dense subgraphs (\textit{constituency divisions}) through overlapping clustering algorithms. 
Consequently, for each subgraph, by jointly considering node attributes and edge links, it generates an elector node (\textit{elector}) with attribute and label learned via the local labeled voter nodes in the original data (\textit{voters}).
Lastly, ELCO connects elector nodes with their corresponding voter nodes, yielding an updated graph.
As pointed out by \citet{top2019ijcai}, for a given graph, higher overall label similarity within the same dense subgraphs can lead to better performance in downstream tasks.
From the generative perspective of graph signals, theoretically and empirically we justify that our newly-generated graph is superior to the original one in terms of the aforementioned similarity.
In addition, we find that this simple pre-processing technique also enhances the class separability of node attributes. For instance, Fig.~\ref{fig:example} illustrates a toy sample of the augmented graph based on the Cora dataset~\citep{data2016icml}, from which we observe that the attributes of $\{N_{1}, N_{2}\}$ are much easier to separate than those of any voter node pair (i.e., $\{n_{i}, n_{j}\} \subset G_{1} \cup G_{2}$).
Therefore, by producing a high-quality augmented training set, ELCO generally renders the performance of subsequent SSL models stronger. We further empirically confirm this claim in experiments.

To validate the practical usefulness of ELCO, we perform extensive evaluations on the SSL benchmark with three standard datasets.
Coupled with two popular models (i.e., Graph Convolutional Network (GCN) and Graph Attention Network (GAT)), our method significantly improves the performance of the base algorithms and consistently outperforms 19 other strong models, including the state-of-the-art.
Moreover, we conduct comprehensive experiments and statistical analyses to provide insights for our method.

In summary, the contribution of this work is four-fold.
\begin{itemize}[noitemsep,topsep=0pt,parsep=0pt,partopsep=0pt]
	\item This is the first study which explores the direction of expanding original graph topology with new nodes and edges to improve training.
	\item By simulating the generation process of graph signals, theoretically and empirically we show that the data augmented by our method has enhanced quality.
	\item This pre-processing technique is entirely agnostic to the input typologies and the subsequent systems, and thus can be coupled with various graph-based SSL models.
	\item In the extensive evaluations of SSL on three standard datasets, our method significantly boosts two popular base algorithms and substantially sets new state-of-the-art scores.
\end{itemize} 

\section{Background}\label{sec:bg} 

For notation purposes, we first formalize the data structure of graph.
Next, we introduce a generative viewpoint for graph signals, which is crucial for obtaining further theoretical insights regarding the proposed algorithm.

\paragraph{Data structure of graph.}

Formally, an attributed graph $G$ (either directed or undirected) can be denoted as $\{V,E,X,Y\}$, where $V=\{v_i\}$ is a set of $|V|$ vertices in $G$, $E\in \mathbb{R}^{n\times n}$ is adjacency relationships between vertices representing the topology of $G$ (i.e., the edge set of $V \times V$), $X\in \mathbb{R}^{n\times d}$ is the $d$-dimensional attribute matrix, and $Y\in \mathbb{R}^{n \times k}$ records the outcome/prediction vectors (with $k$ classes for each vertex label).
$\forall e_{(v_{i},v_{j})} \neq 0, e_{(v_{i},v_{j})} \in E$ denotes there is an edge between $v_{i}$ and $v_{j}$, otherwise $e_{(v_{i},v_{j})} = 0$.
In particular, when $G$ is a directed graph or $E$ contains edge weights, $e_{(v_{i},v_{j})} \neq e_{(v_{j},v_{i})}$.
As a matrix with vertex/node attributes, the $i$th row in $X$ corresponds to the specific attribute of $v_{i}$, which can be regarded as a feature vector with signals from $d$ different channels.
Given $y_{i}$ is a discrete one-hot label vector in matrix $Y$, it also corresponds to vertex $v_{i}$ and the $i$th row of $X$.

\paragraph{A generative viewpoint for graphs.} 
Early works often treat the graph data as a fixed observation(e.g.,~\citet{gcn2017iclr,gat2018iclr}).
However, recent studies show that this perspective has limitations, e.g., in Fig.~\ref{fig:example}, we see that the observation of $P(X|Y)$ may contain non-negligible noise.
Similarly, in practice node attributes and edge links may not correspond to the likelihood of label similarity, i.e., the observation of $P(E|X,Y)$ can be far from the golden distribution, especially in the SSL setting (with only a small number of labeled nodes available)~\citep{nodenoise2020arxiv,ga2019nips}.
For a better real-world approximation, researchers start to view graph data as signals generated from the ground-truth node attributes $X$ and labels $Y$, which can be described by the following factorization of the joint distribution~\citep{lsm2019nips}:
\begin{equation}
P(E,X,Y) = P(E|X,Y)P(Y|X)P(X),
\end{equation}
where the given graph $G$ is treated as a observation of $P(E,X,Y)$, and $P(E|X,Y)$ is the conditional probability of $E$ given $X$ and $Y$.
Because $X$ and $Y$ are not independent, based on the conditional probability formula we have
\begin{equation}
P(E,X,Y) = P(E|X,Y)P(X|Y)P(Y),
\end{equation}
where $X$ can be regarded as the generated data using $Y$ (e.g., in classification task).

\section{Methodology}

\begin{figure*}[tbp]
	\centering
	\includegraphics[width=1.0\textwidth]{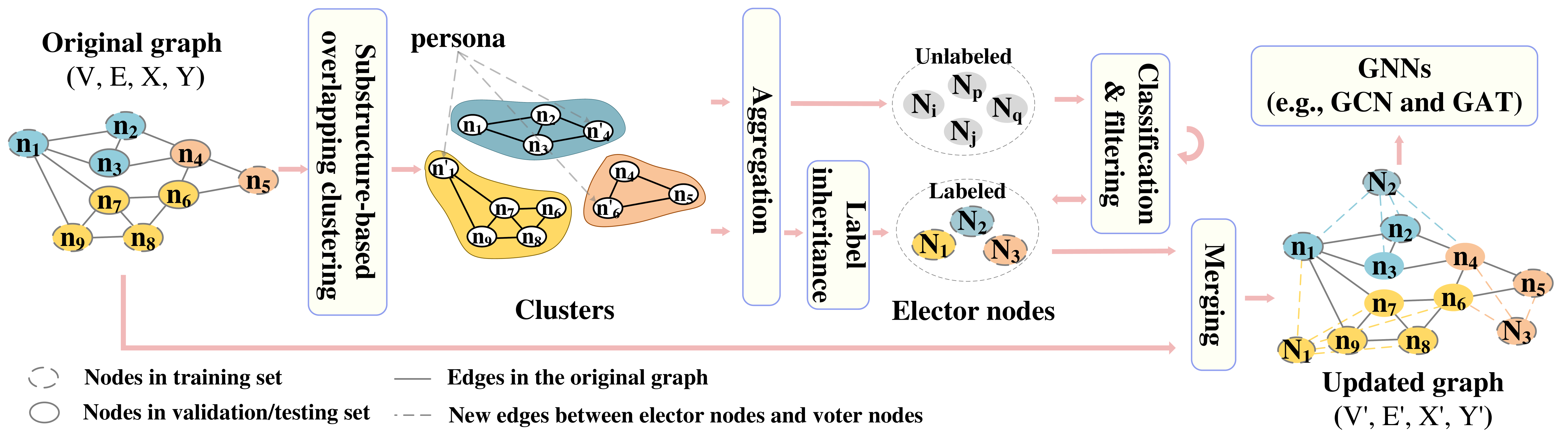}
	\caption{Overview of the ELCO pipeline (please refer to the appendix for pseudo code).}
	\label{fig:overview}
\end{figure*}

By implying the widely-adopted assumption of existing graph-based SSL models, i.e., labels exhibit smoothness along the graph edges~\citep{ga2019nips,gil2020iclr}, \citet{top2019ijcai} propose a criterion to assess training samples, which is highly correlated to the subsequent modeling performance:

\textbf{Criterion $\mathscr{C}$}: \textit{The more nodes in the same dense subgraphs share similar labels, the better the performance of downstream algorithms will achieve.}\label{criterion}

This criterion, which is intuitively evident given the observed presence of graph node communities, has been empirically validated by the experiments of \citet{top2019ijcai}.
Therefore, one of the ultimate objectives of our algorithm boils to: compared with the original graph $G$, the augmented data should satisfy Criterion $\mathscr{C}$ equally well, or even better.

As illustrated in Fig.~\ref{fig:overview}, to achieve this goal, in the first step, we learn to partition the original graph into different dense subgraphs (i.e., clusters).
Next, for each cluster, we automatically generate an elector node, whose attributes can be regarded as the multiple sampling results on attributes of existing voter nodes.
Considering the fact that multiple sampling of distribution can stabilize its posterior probability, compared with voter nodes, elector nodes naturally have better class separability in terms of attributes (which is empirically proven by answering Question~\ref{q1} in experiments).
Consequently, while labels of some elector nodes that can be directly inherited from the voter nodes, the labels of remaining elector nodes can also be roughly determined even with a very simple classifier.
Lastly, since most edges between elector nodes and their corresponding voter nodes can maintain label consistency, the updated graph $G'$ is deemed to have high quality (cf. Criterion $\mathscr{C}$) and has a much larger volume than the original $G$, thus effectively benefiting subsequent graph-based algorithms.
By answering Question~\ref{q2} in experiments, we further present valid evidence to show that in terms of the satisfaction of Criterion $\mathscr{C}$, $G'$ is even better than $G$.

One high-level view of the entire ELCO pipeline is that, it is a continuation of the actual generation process of the given graph $G$.
As discussed in the Background section, $G$ is generated by a random process with $X$ and $Y$ as initial signals.
If $X'$ and $Y'$ are generated from $X$ and $Y$ based on a specific strategy, then $P(X'|Y')P(Y')=P(X|Y)P(Y)$.
Meanwhile, the new edges in $E'$ can be viewed as an extension or self-loop of $E$.
Therefore, in essence, the $G'$ generated by ELCO also uses $X$ and $Y$ as the initial signals and thus can be regarded as the second generation of $G$.

We detail the pipeline of the proposed ELCO as follows.

\subsection{Constituency Division: Substructure-Based Overlapping Clustering}\label{subsec:method_1}

In real-world scenarios, it is quite common that a node belongs to multiple communities (dense subgraphs), e.g., an author publishes a highly impacted paper on machine learning theory, which may get cited by different communities such as computer vision and natural language processing.
Therefore, different from \citet{top2019ijcai} who utilize non-overlapping partitioning approaches, in this paper we identify subgraphs through overlapping clustering, which is a well-studied topic in community detection.
Its workflow of overlapping clustering is two-phased. 
In the first phase, it learns to cluster nodes within local regions. 
To handle nodes belonging to multiple neighborhoods, it will create \textit{personas} for each cluster.
In the second phase, it performs a standard global clustering and re-associates the personas whose sources are the same. 
We denote the resulting set of dense subgraphs as $S'$.

\subsection{Electoral College: High-Level Information Diffusion}\label{subsec:method_2}

From the aforementioned generative perspective of graph, original voter nodes in a cluster $C_{m} \in S'$ become $|C_{m}|$ samples from the golden distribution of attributes. Similarly, the attributes of the corresponding elector node $x_{i}^{c} \in X^{c}$ can be generated through multiple samplings, i.e., aggregating the attributes of voter nodes as
\begin{equation}\label{eq:sample}
x_{i}^{c} = \frac{\sum_{v_{j} \in C_{m}} x_{v_{j}}}{|C_{m}|}, x_{v_{j}} \in X,
\end{equation}
where $X^{c}$ denotes the attribute matrix of all elector nodes, and $Y^{c}$ is for the label matrix, likewise. 
To determine the values of $Y^{c}$, we proceed our discussions case by case.

\paragraph{``Winner takes all".}

For each elector node, if its voter nodes are from the original training set, i.e., have annotated labels, then it can straightforward inherit the dominating (i.e., most numerous) label, such that
\begin{equation}\label{eq:direct}
y_{i}^{c} = [0,\cdots,\underbrace{1}_{j},\cdots,0], n_{j} = \mathrm{max}(n_{0},\cdots,n_{k}),  
\end{equation}
where $n_{p} \ (p=0,\cdots,k)$ is the occurrence of the $p$th kind of label observed in $C_{m}$.
We use $Y^{c}_{obs}$ to denote the label matrix of this category of elector nodes. In practice, we find applying the additional constraint of $n_{j} \geq 2$ can guarantee the reliability of assigned labels.

\paragraph{``Birds of a feather flock together".}

However, only a small portion (e.g., roughly 1/3 in the Cora dataset) of electors nodes can be directly labeled using Eq.~(\ref{eq:direct}).
For other electors nodes, we find that the label propagation paradigm leads to unsatisfactory results, mainly due to two reasons:
on the one hand, real-world graphs (e.g., Cora, Citeseer, and Pubmed~\citep{data2016icml}) are often not fully connected, so many nodes cannot receive the broadcast of training labels; on the other hand, long-range dependencies may bring much noise.
As theoretically explained by \citep{fukunaga2013introduction}, the stability of Eq.~(\ref{eq:sample}) will get strengthened if the number of samplings increases, and $X^{c}$ will thereupon exhibit stronger class separability (of attributes).
This property inspires us to attempt a novel self-learning scheme to handle unlabeled elector nodes.
To begin with, we learn a simple binary classifier based on the already-labeled elector nodes (i.e., the \textit{union} of electors nodes labeled in the last paragraph and those labeled in the previous iterations).
Next, after predicting on-the-fly labels using this classifier, we filter out the elector nodes whose labels are assigned with probability lower than a given ``labeling threshold''.
The above two steps are iteratively performed to produce enough elector nodes with high-quality predicted labels. We use $Y^{c}_{pred}$ for the corresponding label matrix and that $Y^{c}_{rem}$ for that of the remaining unlabeled elector nodes. We fill $Y^{c}_{rem}$ with zeros.

\subsection{Graph Augmentation and Downstream Coupling}\label{subsec:method_3}

By merging high-quality labeled elector nodes into $G$, we obtain the updated graph $G'=\{V', E', X', Y'\}$, where $V' = V \cup V^{c}$, $V^{c}=V^{c}_{obs} \cup V^{c}_{pred} \cup V^{c}_{rem}$ is the set of elector nodes, $X' = X || X^{c}$, $Y' = Y || Y^{c}$, $E' = E \cup\{e_{(v_{i},v_{j}^{c})}\}, v_{i} \in V, v_{j}^{c} \in V^{c}$, and $v_{i} \in C_{j}$.
Apart from the significantly enlarged volume, another outstanding advantage of $G'$ is that, for each dense graph, as elector node becomes the common neighbor of all its voter nodes (i.e., they are linked with new edges), the maximum distance between any two nodes becomes 2, i.e., long-range dependencies get generally shortened.
Empirically, we also observe that $G'$ exhibits a relatively uniform distribution of labels and strong class separability of attributes, both of which can facilitate downstream graph modeling.

Finally, $G'$ can be fed into subsequent graph-based SSL models, with the single aggregation operation for $v_{i}$ at depth $l$ be represented as
\begin{equation}
h_{i}^{l} = \sigma (\sum_{j \in \mathcal{V}_{i} \cup \{n_{i}\}} \alpha_{i,j} W h_{j}^{l-1}),
\end{equation}
where $h_{i}^{l}$ denotes the hidden representation of $v_i$ at the $l$th layer, $\mathcal{V}_{i}$ is the neighbor set of $v_{i}$, $W$ is a learnable linear transformation matrix, $\sigma(\cdot)$ is an element-wise nonlinear activation function, and $\alpha_{i,j}$ is the evaluation parameter set in feature aggregation (e.g., the attention function of GAT).
Stacking multiple such layers with a task-specific layer yields a Graph Neural Network (GNN) with the standard architecture, which can be directly applied in downstream SSL scenarios.

\section{Experiments}\label{sec:exp}

Following previous studies~\citep{data2016icml,gcn2017iclr,gat2018iclr}, we demonstrate the effectiveness of ELCO on the widely-adopted semi-supervised node classification benchmark.
In practice, the amount of labels in graph data is often orders of magnitude smaller than that of all nodes, i.e., $|Y_{obs}| \ll |Y-Y_{obs}|$.
To mitigate this issue, \textbf{graph-based semi-supervised node classification} aims to predict the labels of large-scale nodes with a small training set $(V, E, X, Y_{obs})$.
There exist settings for this task, namely transductive learning and inductive learning, which are different in information visibility.
While the former can fully observe and utilize $X$ in both learning and inference stages, the latter is blocked from partial information (i.e., the features of unlabeled vertices) during the learning stage but are fed with the complete dataset during testing.
In this paper, we focus on the transductive learning.

\subsection{Experimental Setup}

\paragraph{Datasets.}

Our evaluation is based on three datasets (i.e., Cora, Citeseer, and Pubmed~\citep{data2016icml}~\citep{gcn2017iclr}~\citep{gat2018iclr}) which are the \textit{de facto} standards for assessing graph-based SSL algorithms.
They are all sampled via citation networks, where nodes are for research publications and edges for the citation relation.
In Cora and Citeseer, node attributes are represented as bag-of-words, while Pubmed uses TF-IDF weights.
For a fair comparison, we adopt the same training/validation/testing splits as Yang at el.~\citep{data2016icml}, Kipf and Welling~\citep{gcn2017iclr}, and \citet{gat2018iclr}.
Tab.~\ref{table:datasets} shows the detailed statistics of the datasets.

\begin{table}[h]
	\centering
	\scriptsize
	\begin{tabular}{lccccc}
		\toprule
		Data & \#Nodes & \#Edges & \#Features & \#Classes & \#Train/\#Val/\#Test \\
		\midrule
		Cora & 2708 & 5429 & 1433 & 7 & 140/500/1000  \\
		Citeseer & 3327 & 4732 & 3703 & 6 & 120/500/1000 \\
		Pubmed & 19717 & 44388 & 500 & 3 & 60/500/1000 \\
		\bottomrule
	\end{tabular}
	\caption{Summary of the graph datasets.}\label{table:datasets}
\end{table}

\paragraph{Models.}
To justify the generality of ELCO, we respectively integrate it with GCN and GAT as they are the two most popular graph-based SSL methods in the GNN community.
\begin{itemize}
    \item \textbf{GCN}~\citep{gcn2017iclr}, which successfully bridges the gap between spectral and spatial methods. Thanks to its scalability, GCN can efficiently learn node representation by encoding the adjacency matrix and node attributes.
    \item \textbf{GAT}~\citep{gat2018iclr}, which introduces the multi-head self-attention mechanism to achieve the multi-channel information interaction of adjacent nodes.
\end{itemize}

As shown in Tab.~\ref{table:base-results}, we select seven frequently-cited methods and three recently-published approaches as our reference baselines.
In addition, we include six and three variants of GCN and GAT, respectively.
These selected baselines are not the only representative but also very competitive, as some of them claim state-of-the-art performance, such as \citep{gil2020iclr,ga2019nips,adsf2020iclr}.
The results of all the listed baselines are directly duplicated from the corresponding papers, which use the same partition and settings.

\begin{table*}[th]
	\centering
	\small
	\begin{tabular}{c|lccc}
		\toprule
		& Method & Cora & Citeseer & Pubmed \\
		\midrule
		\multirow{10}*{\rotatebox{90}{Reference baselines}} & Gaussian Fields~\citep{gf2013icml} & 68.0 & 45.3 & 63   \\
		& Deep-Semi~\citep{ds2012nn} & 59.0 & 59.6 & 71.7   \\
		& Manifold Reg.~\citep{mr2006imlr} & 59.5 & 60.1 & 70.7   \\
		& Deep-Walk~\citep{dw2014kdd} & 67.2 & 43.2 & 65.3   \\
		& Link-based~\citep{lb2003icml} & 75.1 & 69.1 & 73.9   \\
		& Planetoid~\citep{data2016icml} & 75.7 & 64.7 & 74.4   \\
		& MoNet~\citep{mcnn2017cvpr} & 81.7 & - & 79.0 \\
		& SIG-VAE~\citep{sigva2019nips} & 79.7 & 70.4 & 79.3  \\
		& CurvGN-n~\citep{cgn2020iclr} & 82.7$\pm$0.7 & 72.1$\pm$0.6 & 79.2$\pm$0.5 \\
		& GIL~\citep{gil2020iclr} & 86.2 & 74.1 & 83.1 \\
		\midrule
		\multirow{8}*{\rotatebox{90}{\makecell[c]{GCN-based\\ methods}}} & Chebyshev~\citep{cheby2016nips} & 81.2 & 69.8 & 74.4 \\
		& TAGCN~\citep{tagcn2017axriv} & 83.3 & 72.5 & 79.0 \\
		& TO-GCN~\citep{top2019ijcai} & 83.1 & 72.7 & 79.5 \\
		& DGCN~\citep{dgcn2018www} & 83.5 & 72.6 & 80.0 \\
		& ConfGCN~\citep{confgcn2019aistats} & 82.0$\pm$0.3 & 72.7$\pm$0.7 & 79.5$\pm$0.5  \\
		& LSM\_GCN~\citep{lsm2019nips} & 82.5$\pm$0.2 & 74.4$\pm$0.3 &77.9$\pm$0.4  \\  \rowcolor{gray}
		& GCN~\citep{gcn2017iclr} & 81.5 & 70.3 & 79.0 \\
		& \colorbox{llg}{ELCO-GCN (Ours)} & \colorbox{llg}{85.6$\pm$0.4} & \colorbox{llg}{75.7$\pm$0.3} & \colorbox{llg}{83.7$\pm$0.3} \\
		\midrule
		\multirow{5}*{\rotatebox{90}{\makecell[c]{GAT-based\\ methods}}} 
		& LSM\_GAT~\citep{lsm2019nips} & 82.9$\pm$0.3 & 73.1$\pm$0.5 & 77.6$\pm$0.7 \\
		& GAT$_{128}$+GAM*~\citep{ga2019nips}\footnotemark & 85.0 & 73.6 & - \\
		& ADSF-RWR~\citep{adsf2020iclr} & 85.4$\pm$0.3 & 74.3$\pm$0.4 & 81.2$\pm$0.3 \\ 
		& GAT~\citep{gat2018iclr} & 83.0$\pm$0.7 & 72.5$\pm$0.7 & 79.0$\pm$0.3 \\
		& \colorbox{llg}{ELCO-GAT (Ours)} & \colorbox{llg}{\textbf{87.6$\pm$0.5}} & \colorbox{llg}{\textbf{76.7$\pm$0.4}} & \colorbox{llg}{\textbf{84.3$\pm$0.3}} \\
		\bottomrule
	\end{tabular}
	\caption{Accuracy (\%) of the node classification benchmark. The highest performance per dataset is highlighted in \textbf{bold}. The $\pm$ error bar denotes the standard deviation in 10 independent trials. `-' means the corresponding value has not been published in the original paper.}
	\label{table:base-results}
\end{table*}
\footnotetext{In particular, GAT$_{128}$+GAM uses 128 hidden units, which is more than the original GAT (16).}

\paragraph{Parameters.}

In practice, we find that our proposed ELCO is robust towards configuration variations (Our code is available in supplementary material).
Therefore, we exploit the most straightforward setting for parameters without much fine-tuning.
To be specific, for overlapping clustering we select the robust Ego-Splitting algorithm~\citep{egosplit2017kdd,karateclub2020}), with the resolution set at 1.0\footnote{ELCO consistently yields positive results with different overlapping clustering methods and setups. Due to space limitations, please see our technical appendix for more details.}.
In the high-level information diffusion step, we utilize the simple GBDT~\citep{gbdt2001jstor} as our classifier, with a learning rate at 0.25, max depth at 3, and other parameters selected as default.
To ensure the quality of generated labels, we set the number of iterative diffusions and the labeling threshold at 10 and 0.99, respectively. 
During all experiments, we terminate the training when the verification accuracy no longer increases for 2K iterations.
Test scores based on the models with the best verification performance are reported, and all the hyper-parameters used are listed. 

\subsection{Main Results}

Tab.~\ref{table:base-results} reports the results of our baseline methods, base algorithms (GCN and GAT), and ELCO-enhanced models (ELCO-GCN and ELCO-GAT).
To reduce randomness, we run each model for 10 independent trials and calculate the average score and standard deviation.
On all the three datasets, ELCO-GAT consistently sets new state-of-the-art performance, with margins of 0.6\% to 3.3\% compared with the best baselines which are not coupled with ELCO.
Before being stacked with ELCO, the base GCN is 1.2\% inferior to GAT on average; after the data augmentation, ELCO-GCN still falls behind ELCO-GAT.
However, compared with other baseline methods, ELCO-GCN ranks second on Citeseer and Pubmed and third on Cora, exhibiting strong competitiveness.
Please note that neither GCN nor GAT achieves outstanding accuracies compared with their strong counterparts: more concretely, even the original GAT cannot rank within the top three (with ELCO-enhanced models excluded) on any dataset. This fact emphasizes the substantial effectiveness of ELCO.

When calculating the specific accuracy enhancement brought by ELCO, we witness very significant 4.1\% to 5.4\% and 4.2\%  to 4.7\% increases for GCN and GAT, respectively.
Put these increments in the context: among all the other GCN-based approaches, the range of performance gain over GCN is -4.6\% to 4.1\%; among all the other GAT-based ones, it is -1.4\% to 2.4\% over GAT.
It is therefore recommended that ELCO be adopted as a standard by graph-based SSL pipelines.

\subsection{Further Discussion}\label{sec:further_dicussion}
To obtain more insights for our proposed methods, we perform experiments to investigate the following research questions:

\newcounter{question}
\noindent\textbf{Question 1}{\refstepcounter{question}\label{q1}}: Do elector nodes have better attribute class separability than voter nodes?
\\ \noindent\textbf{Question 2}{\refstepcounter{question}\label{q2}}: Is $G'$ superior to $G$ in terms of their satisfaction of Criterion $\mathscr{C}$? 

\begin{figure}[tbp]
	\centering
	\includegraphics[width=0.48\textwidth]{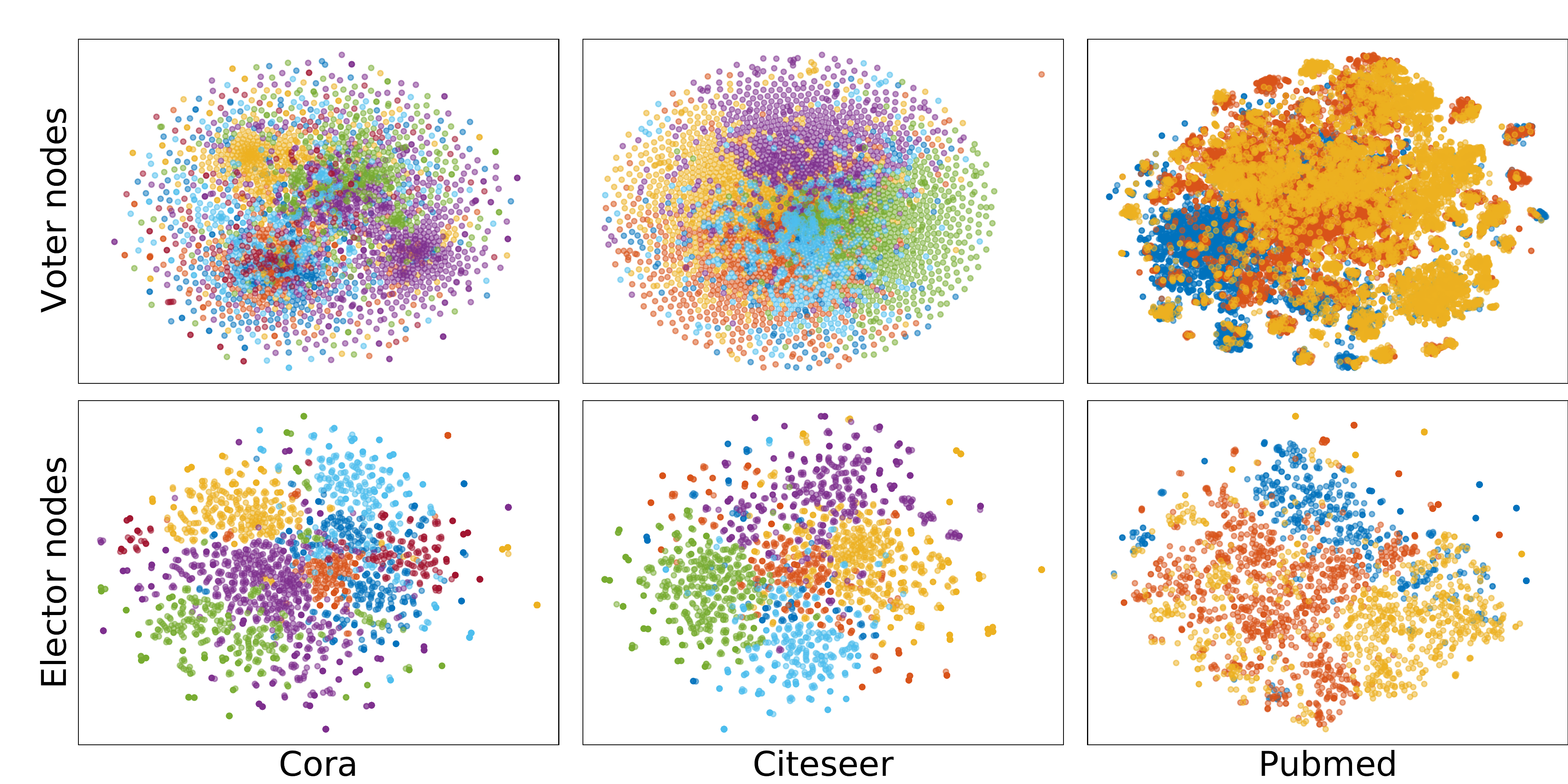}
	\caption{The visualization results of the attributes of the original nodes and the elector nodes after the dimension reduction by using t-sne.
		The denser the nodes with the same label and the more scattered the nodes with different labels, the stronger the class separability of attributes.}
	\label{fig:attributes}
\end{figure}

\begin{table}[t]
	\centering
	\begin{subtable}{.45\textwidth}
		\centering
		\setlength{\belowcaptionskip}{8pt}
		\small
		\begin{tabular}{lccc}
			\toprule
			& Cora & Citeseer & Pubmed  \\
			\midrule	 
			voter node & 28.23 & 32.46 & 13.14   \\
			elector node & \textbf{4.41} & \textbf{6.37} & \textbf{5.65}   \\
			\bottomrule
		\end{tabular}
		\caption{Error Rate of Linear Classifier (L2).}\label{table:measure}
	\end{subtable}
	\hspace{\fill}
	\begin{subtable}{.45\textwidth}
		\centering
		\setlength{\belowcaptionskip}{8pt}
		\small
		\begin{tabular}{lccc}
			\toprule
			& Cora & Citeseer & Pubmed  \\
			\midrule	 
			voter node & 76.63 & 74.45 & 85.43  \\
			elector node & \textbf{91.51} & \textbf{88.38} & \textbf{99.97}   \\
			\bottomrule
		\end{tabular}
		\caption{Classification accuracy (\%) of GBDT.}\label{table:iterative_learning}
	\end{subtable}\caption{Results of attribute class separability comparisons.}
\end{table}

To begin with, in Fig.~\ref{fig:attributes}, we visualize (original) voter nodes and (generated) elector nodes with attributes as axes.
While the former seem more crisscross, the latter exhibit clearer ``community borders", demonstrating that elector nodes have better class separability of attributes.
For more rigours comparisons, we leverage two widely-adopted metrics.
Firstly, for voter and elector nodes, we measure the Error Rate of Linear Classifier (L2), i.e., SVM, which is the direct estimation of separability~\citep{l22019cs} (lower L2 means better separability).
As shown in Tab.~\ref{table:measure}, on all datasets, elector nodes yield significantly-lower L2, indicating their overall separability is much better.
In addition, we investigate a task-driven metric, i.e., to see if node attributes can be precisely classified with a simple model.
In Tab.~\ref{table:iterative_learning}, we observe an average accuracy gap of 14.5\%, meaning that elector nodes are easier to get classified with attribute information.
Note that the classifier chosen is GBDT, which is consistent with our implementation of ELCO. 
Consequently, results in Tab.~\ref{table:iterative_learning} also shows that the generated elector nodes are of high quality. 
The above empirically gives a positive answer for Question~\ref{q1}.

\begin{figure}[t]
	\centering
	\includegraphics[width=0.48\textwidth]{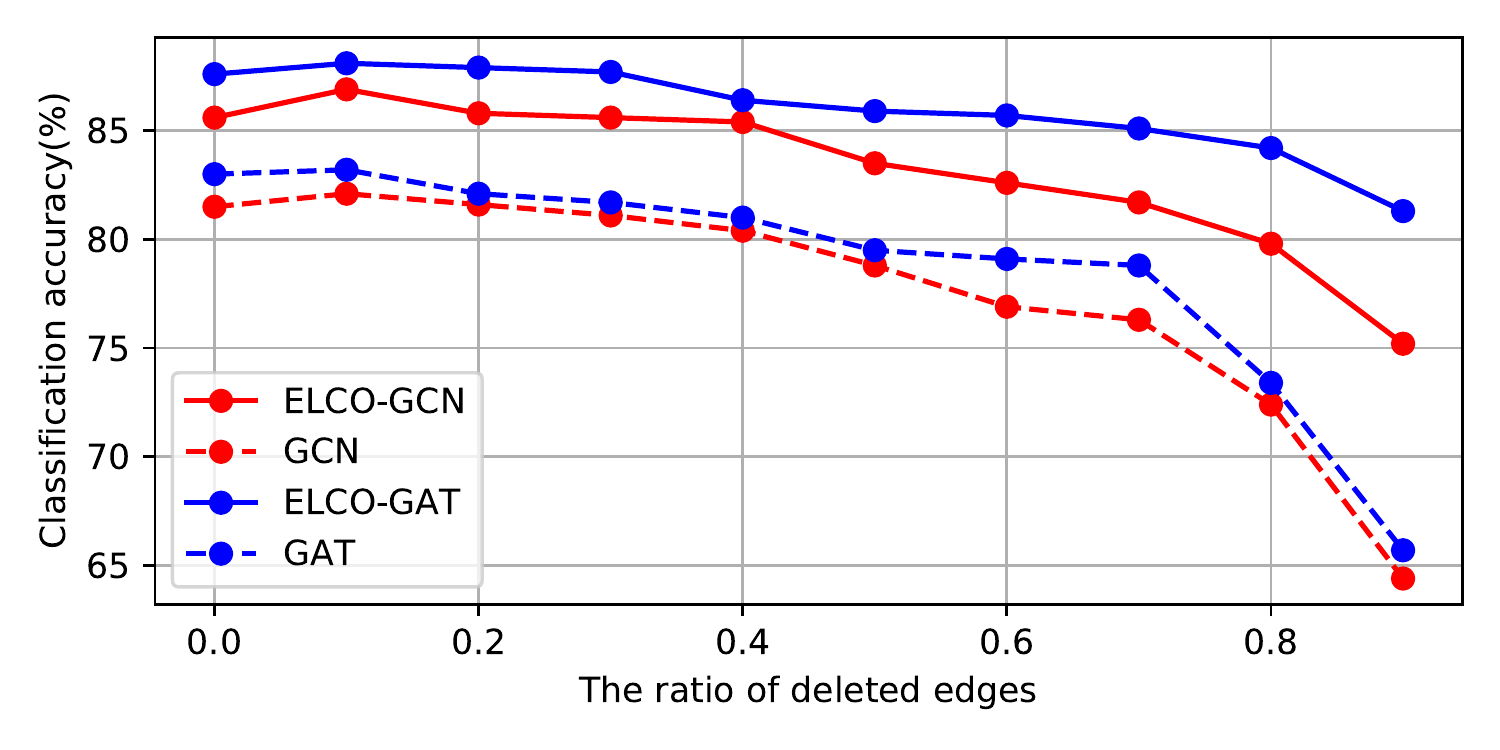}
	\caption{
		Relationship between edge sparsity and model performance in Cora.
	}
	\label{fig:removeedge}
\end{figure}

\begin{figure*}[t]
	\centering
	\includegraphics[width=1.0\textwidth]{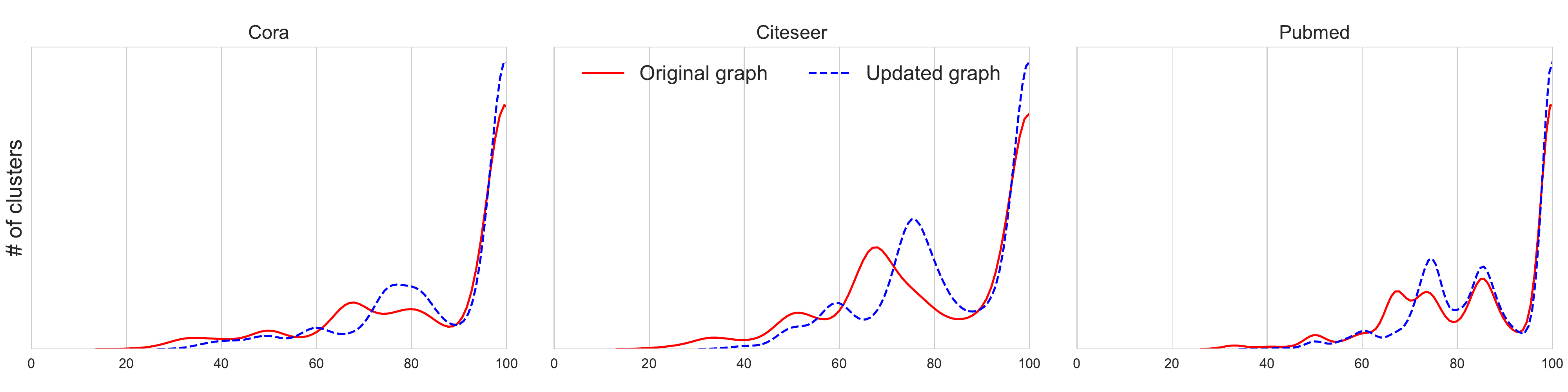}
	\caption{Distribution of per cluster dominating label proportion.
		X-axes denote the proportion of nodes which have the dominating label in each cluster.
	}
	\label{fig:label_proportion}
\end{figure*}

Next, to assess the label similarity within a dense subgraph, we design a straightforward experiment, which counts the percentage of dominating labels in each cluster partitioned by ELCO).
We parallel to evaluate the original $G$ and the updated $G'$ and plot the occurrence densities of dominating label proportion in Fig.~\ref{fig:label_proportion}.
On all datasets tested, compared with the curves in $G$, those in $G'$ have higher peaks and are more right-gathered, implying that a larger proportion of nodes with subgraphs share the same labels, e.g., for the Cora dataset, more than 50\% of the clusters have 100\% of the labels to be same (i.e., the proportion of dominating labels is 1). We thereupon verify that ELCO enhances the label similarity, i.e., reply yes to Question~\ref{q2}.

In addition, the density of edges in the graph has been shown to affect the performance of downstream tasks (e.g., community detection, label propagation), so the performance of the model in the sparse graph needs to be discussed.
To test the performance of ELCO in the different denseness of the edges in the graph, we randomly delete some edges in the datasets for comparison.
The experimental results are shown in Fig.~\ref{fig:removeedge}. 
As part of the edges are removed, the performance of the model is improved briefly (verified in~\citep{dropedge2020iclr}) and then degraded sharply after most of the edges are removed.
Compared with the original methods, ELCO is more robust and accurate in the initial stage, and still better than the original methods after removing most of the edges.
While eliminating edges destroys part of the community structure, ELCO can make up for lost information by relying on the rest of the structure to supplement it.

\subsection{Ablation Studies on the Self-learning Module}
To see how the composition of nodes fed into the simple classifier (i.e., GBDT) affects the prediction accuracy when generating $Y^{c}_{pred}$, we conduct ablation studies on various versions of samples:
already-labeled elector nodes (i.e., our implemented version), annotated voter nodes in $G$, and the mixture of both. 

\definecolor{dark_green}{rgb}{0, 0.5, 0}
\begin{figure*}[tbp]
	\centering
	\includegraphics[width=1.0\textwidth, trim={.6cm .0cm 1.cm .0cm},clip]{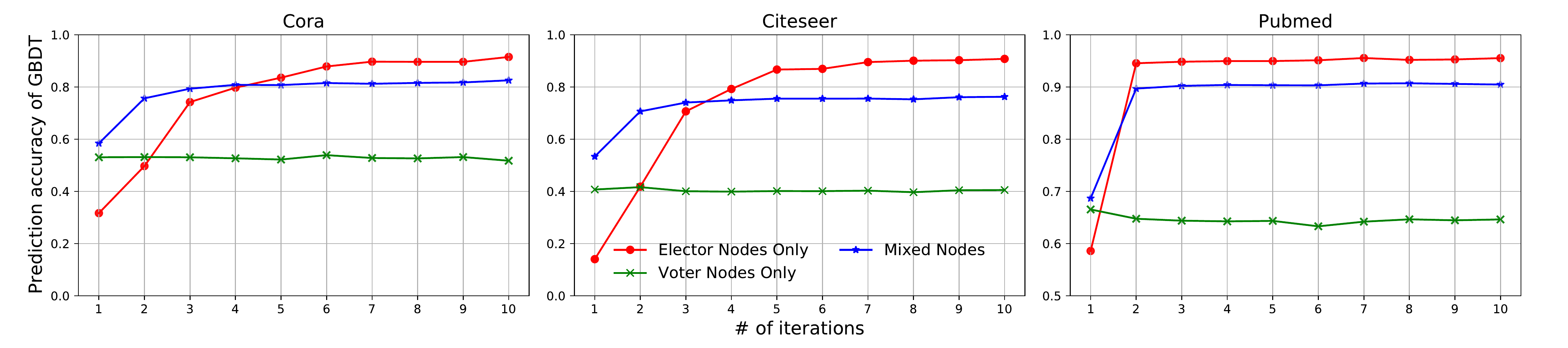}
	\caption{The prediction accuracy growth of GBDT with different training node compositions.
		}
	\label{fig:self-learn}
\end{figure*}

As illustrated in Fig.~\ref{fig:self-learn}, training with voter nodes only cannot even converge (i.e., the binary classification accuracy is always around 0.5 and non-growing).
We identify the unsatisfactory attribute class separability as the leading cause: it makes training a simple classifier unrealistic.
If elector nodes are used for training GBDT, we can see that a large number of high-quality labeled nodes can get harvested through iterative training, thanks to the strong class separability.
If we mix two types of nodes, the self-learning process can achieve convergence. However, the final prediction accuracy still significantly falls behind our implemented setting, i.e., using elector nodes only.

\section{Related Work}\label{sec:rw}

\paragraph{Generative graph-based SSL models.} Due to the inherent uncertainty of real-world graphs~\citep{muller2012neural}, the community has witnessed an increasing interest in analyzing graphs with generative models.
Specifically, some recent studies attempt to weaken the uncertainty by encapsulating all relevant information of nodes (i.e., generated samples) and realize label propagation.
For example, \citet{lsm2019nips} exploit scalable variational inference to approximate the Bayesian posterior of the joint distribution of node features, predictions, and graph structure.
Similarly, based on encapsulating attributes, paths, and local graph structures, \citet{gil2020iclr} propose a graph inference learning framework to model node labels topologically.
In addition, \citet{cgn2020iclr} leverage discrete graph curvature to measure to what extent the neighborhoods of a node pair are structurally related.

\paragraph{Topological refinement algorithms.} On the other hand, some authors attempt to reduce the impact of the aforesaid uncertainty by adjusting the topology of graph.
DropEdge~\citep{dropedge2020iclr} randomly removes edges to prevent excessive smoothing.
The model of \citet{mixhop2019icml} mixes the features of multi-hop neighbors by short-circuiting distant nodes.
\citet{top2019ijcai} adjust both inter-community and intra-community edges to optimize the graph topology.
\citet{align2020arxiv} maximize consistency for aggregate information by aligning networks at both topological and semantic levels.
\citet{nodenoise2020arxiv} incorporate a robust norm feature learning mechanism with graph convolution for SSL with constraints.

To our knowledge, ELCO is the first approach to bridge the gap between graph generative models and topological refinement algorithms, which improves the original graph by adding high-quality nodes generated through joint modeling structure and attribute signals.

\section{Conclusion}

In this paper, we propose a simple yet effective ELCO framework, which boosts the performance of graph-based SSL models by augmenting training resources.
As the first attempt to expand the existing typology, our pipeline can be regarded as a continued graph generation process based on the input information.
Aiming to strengthen the label similarity within dense subgraphs, ELCO generates high-quality nodes, which also exhibit refined class separability of attributes.
Results of extensive evaluations indicate that this generic pre-processing technique can dramatically enhance the base algorithms and further outperform state-of-the-art baselines.
Followup experiments and analyses present more insights regarding the superiority of ELCO.
In the future, we will test ELCO in more setups, as well as explore other graph augmentation strategies.

\section*{Broad Societal Implications}
The ELCO framework significantly boosts the performance of graph-based semi-supervised models, thus benefiting real-world applications which process graph data, e.g., social networks, sensor networks, and molecular structures. 
Apart from the node classification task which is tested, ELCO may be further applied in a wider spectrum of tasks, including link prediction, data completion, and graph generation.

Apart from a generic augmentation algorithm for graph data, our method can also be regarded as a novel approach to exploiting the existing internal dependencies in low-resource scenarios. Thus, if a specific task from other research fields (e.g., natural language processing, computer vision, data mining, etc.) involves data with characteristics of graph signals, e.g., smoothness/similarity of features within neighboring regions, then ELCO can be possibly adopted.

Please also be aware of some known risks and limitations of our framework. Firstly, when the given graph is too sparse or its internal connections are close to saturation, ELCO may fail to provide satisfactory results, i.e., the augmented data is too few or has low quality.
Besides, without explicit mechanisms to handle bias originally introduced by the input, ELCO may yield biased output.  
Lastly, since the generation of new nodes depends on dense subgraphs, an excessive number of such structures may bring considerable computational overhead.

We would encourage researchers to explore further applications of our method.
To mitigate the aforementioned risks and limitations and improve the real-world usability of ELCO, we also welcome all kinds of improvements and enhancements from any research field.

\bibliography{ref}

\end{document}